\documentclass[%
reprint,
superscriptaddress,
 amsmath,amssymb,
 aps,
]{revtex4-1}

\usepackage{graphicx}
\usepackage{dcolumn}
\usepackage{bm}
\usepackage[colorlinks, linkcolor=red, anchorcolor=blue,urlcolor=blue, citecolor=blue]{hyperref}

\newcommand{\beq}{\begin{equation}}
\newcommand{\eeq}{\end{equation}}
\newcommand{\beqa}{\begin{eqnarray}}
\newcommand{\eeqa}{\end{eqnarray}}

\begin{document}


\title{Inverse engineering for fast transport and spin control of spin-orbit-coupled Bose-Einstein condensates in moving harmonic traps}

\author{Xi Chen}
\email[]{xchen@shu.edu.cn}
\affiliation{Department of Physics, Shanghai University, 200444 Shanghai, People's Republic of China}%

\author{Ruan-Lei Jiang}
\affiliation{Department of Physics, Shanghai University, 200444 Shanghai, People's Republic of China}%

\author{Jing Li}
\affiliation{Department of Physics, Shanghai University, 200444 Shanghai, People's Republic of China}%

\author{Yue Ban}
\email[]{yban@shu.edu.cn}
\affiliation{Department of Electronic Information Materials, Shanghai University, 200444 Shanghai, People's Republic of China}%
\affiliation{Instituto de Ciencia de Material de Madrid, CSIC, 28049 Madrid, Spain}

\author{E. Ya. Sherman}
\affiliation{Department of Physical Chemistry, Universidad del Pa\'{\i}s Vasco UPV-EHU, 48080,
Bilbao, Spain}
\affiliation{IKERBASQUE Basque Foundation for Science, Bilbao, Spain}

\date{\today}

\begin{abstract}
We investigate fast transport and spin manipulation of tunable spin-orbit-coupled Bose-Einstein
condensates in a moving harmonic trap. Motivated by the concept of ``shortcuts to adiabaticity",
we design inversely the time-dependent trap position and spin-orbit coupling strength. By choosing
appropriate boundary conditions we obtain fast transport and spin flip simultaneously.
The non-adiabatic transport and relevant spin dynamics are illustrated with numerical examples,
and compared with the adiabatic transport with constant spin-orbit-coupling strength and velocity. Moreover, the influence of nonlinearity
induced by interatomic interaction is discussed in terms of the Gross-Pitaevskii approach,
showing the robustness of the proposed protocols.
With the state-of-the-art experiments, such inverse engineering technique paves the way for coherent control
of spin-orbit-coupled Bose-Einstein condensates in harmonic traps.

\end{abstract}

\maketitle

\section{INTRODUCTION}

Spin-orbit-coupling (SOC), linking a quantum particle's momentum to its spin, is a fundamental effect in
solid-states spintronics \cite{Review1}.
In recent years, the experimental breakthrough, realizing a synthetic SOC for (pseudo) spin-1/2 bosonic \cite{Spielman}
and fermions \cite{Zwierlein,ZhangFermi}, has provided a platform for
quantum simulation of exotic states in condensed matter physics and a flexible tool for manipulating cold atoms,
see review \cite{SpielmanRev,ZhaiH}.
In particular, the static and dynamical properties, relevant to the SOC effects, have been extensively investigated in
such atomic systems \cite{MuMW,LiuWM,Engels,LZSOCBEC,SpielmanLZ,WangPRA,Dimitri,ZhangCWPRA,Ji},
which open new possibility to probe or control quantum spin dynamics such as spin relaxation, Zitterbewegung,
spin resonance, and the spin-Hall effect.

Cold atoms are comprehensively stored and manipulated in traps formed by designed electromagnetic
field configurations, with fundamental interest and potential applications
in atom interferometry, metrology or quantum information processing. Very often, a transport of neutral
or ionized cold atoms and Bose-Einstein condensates (BECs)
to appropriate location without any excitation and losses \cite{Hansel,Esslinger,Ketterle,DavidEPL,Donner}
is demanded.
However, most transport processes require long time, satisfying the slow adiabatic criteria,
which could be problematic due to the noise and decoherence. An alternative way out is to apply the
concept of ``shortcuts to adiabaticity" \cite{STA} to reach the same results at a relatively short time.
Among all shortcut techniques, the inverse engineering method, based on Lewis-Riesenfeld invariant and
corresponding dynamical modes, can be also applicable to achieve fast non-adiabatic but reliably
controllable transport \cite{ErikPRA,ErikBEC,ZhangQ,Lu14,David2}, or expansion \cite{Muga,ChenPRL} in harmonic
traps and spin control in (effective) two-level systems \cite{YuePRL}. More interestingly, the spin and motional
states can be precisely and simultaneously controlled by such inverse engineering method in a Morse potential with SOC \cite{MorseYue}.
In such static potential trap, the direction and magnitude of the synthetic SOC field are chosen as tunable parameters in effective two level system to manipulate spin states, and the position transfer does result from the effect of SOC on the orbital motion, rather than the modulation of the trap center.


In this paper, we propose a method for controlling spin dynamics and orbital motion of BECs in moving harmonic
traps with Raman-induced SOC. Similarly to electrons in quantum dots \cite{ZhenyaPRB,RamsakNJP,RamsakPRL,JQYouPRL},
cold atoms are confined in harmonic traps, and the spin state and orbital motion can be controllable by time-dependent SOC.
The exact wave function of atoms trapped in a moving harmonic trap in presence of time-dependent SOC can be solved analytically \cite{RamsakNJP,RamsakPRL},
to demonstrate the controllability of spin state and orbital motion. Instead of (non-adiabatic) cyclic evolution in Ref. \cite{RamsakPRL},
we apply inverse engineering approach to design the position of moving trap and time-dependent strength of SOC, in order to transport cold atoms, arriving at
appropriate location with spin flip simultaneously.
The non-adiabatic transport and relevant spin dynamics are illustrated with numerical examples. As compared with
adiabatic transport with constant SOC strength and velocity,
we illustrate that inverse engineering provides more flexibility to manipulate the cold atoms and spin-orbit
qubits in a fast and robust way.
By extension, we also discuss the SOC BECs in the presence of interaction between the atoms and show the stability
against the effects of nonlinearity. The results presented below thereby may acquire wide applications.

\section{General equation for orbital and spin motion}

\begin{figure}[tbp]
\scalebox{0.40}[0.40]{\includegraphics{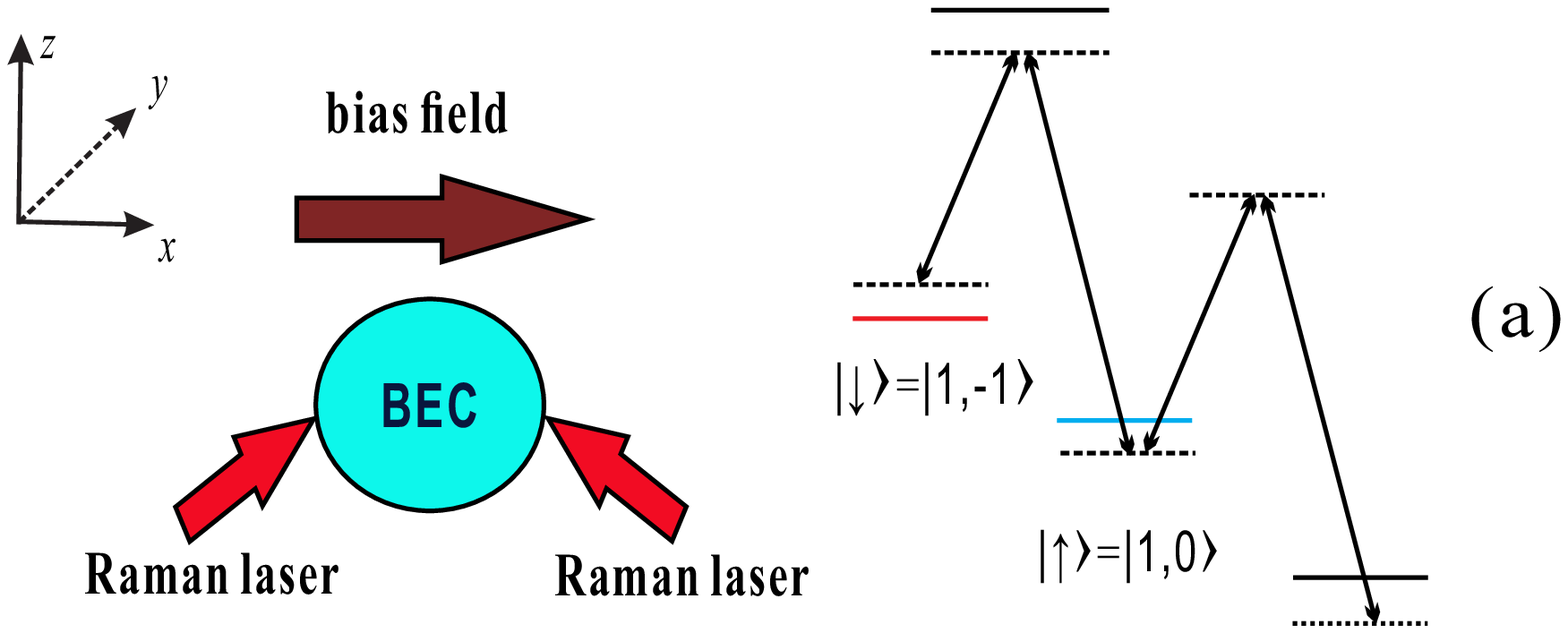}}
\\
\scalebox{0.40}[0.40]{\includegraphics{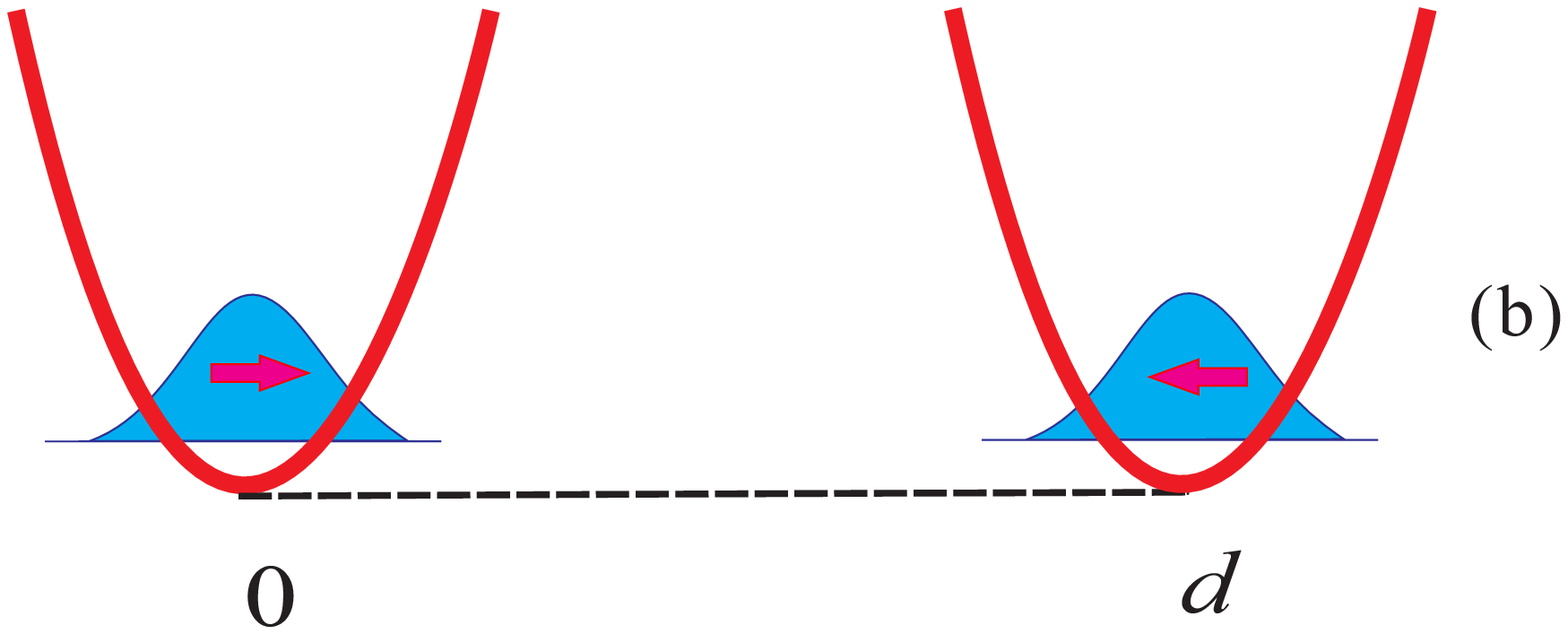}}
\caption{(a) Experimental scheme for realizing synthetic SOC BECs and the level diagram.
A homogeneous magnetic field prepares the ground state spin-polarized along the $x$-direction.
Two Raman lasers are used to couple the spin-up and spin-down states. (b) Schematic diagram of the BECs transport
in an effective one-dimensional harmonic potential from $x_{0}=0$ to $x_{0}=d$.  }
\label{figscheme}
\end{figure}

Our starting point is the BECs in a one-dimensional (along the $x-$axis) harmonic potential with SOC, see Fig. \ref{figscheme},
where a cloud of ultracold $^{87} \mbox {Rb}$ atoms is strongly confined in the $y$-$z$ plane.
Internal hyperfine ground states $|\uparrow\rangle = |F=1, m_f=0\rangle$ and $|\downarrow \rangle =|F=1, m_f=-1\rangle$
coupled by two Raman lasers can be identified here as (pseudo-)spins.
The dynamics in a moving harmonic potential is governed by the following Hamiltonian
\begin{equation}
\label{Hamiltonian}
H=\frac{p^{2}}{2m}+\frac{1}{2}m\omega^{2}[x-x_{0}(t)]^{2}+\alpha(t)p\sigma_{z},
\end{equation}
where $p$ and $x$ are the momentum and position operators, respectively, $m$ is the particle mass, $\omega$
is the time-independent potential frequency, and $\sigma_{z}$ is the corresponding
$2\times2$ Pauli matrix.
Here $x_{0}(t)$ is the time-dependent trap position,
which can be tuned, for instance, by changing Gaussian beam waist center \cite{Donner}.
The parameter $\alpha(t)$ is a controllable SOC strength,
adjusted by the geometry of two Raman lasers \cite{Spielman}, see Fig. \ref{figscheme} (a).
Here we neglect the Zeeman term, $\Delta \sigma_x$, since the magnetic field is supposed to be
switched off after the initial spin state is prepared.

We present the $|\Psi(x,t)\rangle\equiv[\Psi_{\uparrow}(x,t),\Psi_{\downarrow}(x,t)]^{\rm T}$ solution of the time-dependent
Schr\"{o}dinger equation with Hamiltonian (\ref{Hamiltonian}) in the form:
\begin{equation}
\label{solution}
|\Psi(x,t)\rangle=e^{-iE t/ \hbar} U(t)|\psi(x)\rangle|\chi_s\rangle,
\end{equation}
where $E$ is the eigenvalue, $U(t)$ is a unitary transformation, $|\chi_s\rangle$ is a
spinor with spin $s$, $|\psi(x)\rangle$ stands for the eigenfunction of the stationary harmonic oscillator
\beq
\label{harmonic}
\left[\frac{p^2}{2m}+\frac{1}{2} m \omega^2 x^2 \right]\psi(x)=E \psi(x).
\eeq
Following the approach of Refs. \cite{RamsakNJP,RamsakPRL}, we introduce the unitary operator, $U(t)=U_s(t) U_o(t)$, with the spin and orbit parts,
\beqa
\label{U}
\nonumber
U_s(t)&=&e^{-i\phi_{\alpha}(t)} e^{-i\phi(t)\sigma_{z}} e^{-ima_{c}(t)x\sigma_{z}/\hbar} e^{-i\dot{a}_{c}(t)p\sigma_{z}/(\hbar\omega^{2})}\\
U_o(t)&=& e^{-i\phi_{x_{0}}(t)} e^{-ix_{c}(t)p/\hbar} e^{im\dot{x}_{c}(t)x/\hbar},
\eeqa
where the upper dot represents time derivative, and we shall seek for the yet unknown phase factors which determine their time-dependence.
Here two action phase factors:
\beqa
\phi_{\alpha}(t)=-\frac{1}{\hbar}\int_0^t d\tau \mathcal{L}_{\alpha}(\tau),
\\
\phi_{x_{0}}(t)=-\frac{1}{\hbar}\int_0^t d \tau  \mathcal{L}_{x_{0}}(\tau),
\eeqa
are expressed with the Lagrangians for classical mechanics:
\beqa
\mathcal{L}_{\alpha}(t)&=& \frac{1}{2\omega^{2}} m \dot{a}_{c}^{2}(t)- \frac{1}{2} m a_{c}^{2}(t)+m a_{c}(t)\alpha(t),
\\
\mathcal{L}_{x_{0}}(t) &=& \frac{1}{2} m \dot{x}_{c}^{2}(t)- \frac{1}{2} m\omega^{2}[x_{c}(t)-x_{0}(t)]^{2}.
\eeqa
To guarantee that Eq. (\ref{solution}) is the exact solution, we introduce two auxiliary functions, $x_{c}(t)$ and $a_{c}(t),$
satisfying equations
\beqa
\label{eqxct}
\ddot x_{c}(t)+\omega^{2}[x_{c}(t)-x_{0}(t)] &=& 0,
\\
\label{eqact}
\ddot a_{c}(t)+\omega^{2}[a_{c}(t)-\alpha(t)] &=& 0,
\eeqa
which describe the center-of-mass position of the BECs and its spin precession, respectively. Also the phase
factor, standing for the spin rotation along the $z$-direction, couples these two parameters as
follows
\beq
\label{phase}
\phi_{\sigma}(t)=-\frac{m}{\hbar}\int_0^t \dot{a}_{c}(\tau)x_{0}(\tau)\,d\tau.
\eeq
Without SOC, $\alpha(t)=0$, the second auxiliary Eq. (\ref{eqact}) becomes trivial, thus the problem is reduced to
the previous transport design \cite{ErikPRA,ErikBEC} by using inverse engineering.
As a matter of fact, when the interaction between atoms is negligible, the wave function of
in Eq. (\ref{solution}) is nothing but the transport
modes based on Lewis-Riesenfeld dynamical invariant, with the eigenvalue $E_n=(n+1/2)\hbar \omega$ and eigenstate  $|\psi(x)\rangle$
of a stationary harmonic trap \cite{ErikPRA}, see Eq. (\ref{harmonic}). Additionally,
the Hamiltonian (\ref{Hamiltonian}) resembles the one for electron in a moving quantum dot with time-dependent SOC \cite{RamsakNJP,RamsakPRL}.
Here we shall concentrate on the transport of BECs with tunable SOC.

In what follows we shall develop the inverse engineering method for transporting BECs rapidly and
flipping the spin simultaneously by using Eqs. (\ref{eqxct})-(\ref{phase}).
This is substantially different from the holonomic transformation of spin-orbit qubits
\cite{RamsakPRL}, where the controllable parameters are periodically modulated, thus the spin rotation is determined by the non-adiabatic Aharonov-Anandan phase for cyclic time evolution.
The strategy presented here is that we first consider the position of moving potential $x_0 (t)$ and the time-dependent SOC strength $\alpha(t)$ as free
controllable parameters,  and then design them inversely based on the solutions of $x_c(t)$ and $a_c(t)$
satisfying the appropriate boundary conditions. Thus, the spin rotation and orbital motion can be simultaneously manipulated as we wish.
To illustrate the technique, we begin with the linear transport of the BECs with time-dependent SOC, neglecting interatomic interaction.
Later, we will check the influence of nonlinearity resulting from interaction between the atoms by a numerical simulation.
The designed shortcut protocol will be compared with the adiabatic transport with constant velocity and SOC strength, showing the
advantages of inverse engineering proposed here.

\section{INVERSE ENGINEERING}

To design the potential position and SOC strength inversely, we focus on Eqs. (\ref{eqxct}) and (\ref{eqact})
by choosing appropriate boundary conditions.
Suppose that the potential minimum moved from $x_{0}(0)=0$ at initial time $t=0$ to $x_{0}(t_{f})=d$ within a time interval $t_{f}$.
To guarantee the transport without a final excitation, we set the following boundary
conditions \cite{ErikPRA,ErikBEC}:
\begin{eqnarray}
\label{conditionxc1}
  x_{c}(0)&=&0,  \dot{x}_{c}(0)=0, \ddot{x}_{c}(0)=0, \\
\label{conditionxc2}
  x_{c}(t_{f})&=&d,  \dot{x}_{c}(t_{f})=0, \ddot{x}_{c}(t_{f})=0.
\end{eqnarray}
These boundary conditions can be satisfied by an infinite set of functions. For simplicity, we choose
a flexible polynomial ansatz in the form, $x_{c}(t)=\sum^{5}_{i=0}b_{i}t^{i}.$ Finally, we
can obtain the center-of-mass position of the cold atoms $x_{c}(t)$ as:
\beq
x_{c}(t)=d(10s^{3}-15 s^{4}+6 s^{5}),
\eeq
with $s\equiv\,t/t_f$, which provides the desired $x_{0}(t)$ obtained from Eq. (\ref{eqxct}),
namely, $x_0 (t) = x_c (t) + \ddot{x}_c (t) /\omega^2$.

Now we can fulfill the task of the spin flipping without excitation of the orbital motion.
To ensure that the effects of $a_c(t)$ and $\dot{a}_c(t)$ in
the unitary transformation (\ref{U}) vanish at the initial and final time,
we set the boundary conditions:
\begin{eqnarray}
\label{conditionac1}
a_{c}(0) &=&a_{c}(t_{f})=0, \\
\label{conditionac2}
 \dot{a}_{c}(0) &=& \dot{a}_{c}(t_{f})=0,\\
\label{conditionac3}
 \ddot{a}_c (0) &=& \ddot{a}_{c} (t_{f})=0.
\end{eqnarray}
Here the boundary conditions for the second derivative avoid the abrupt changes in the SOC strength at the edges, $t=0$ and $t_f$,
which could be implemented by switching on/off the Raman laser.
In addition, the phase factor of the spin rotation, $\exp[-i\phi_{\sigma}(t)\sigma_z]$, acts on
the initial eigenstate of $\sigma_x$ ($[1/\sqrt{2},1/\sqrt{2}]^{\rm T}$)
in the $\sigma_{z}$-representation. As a result, when $\phi_{\sigma}(t_{f})=\pi/2$, the spin state
changes to $[1/\sqrt{2},-1/\sqrt{2}]^{\rm T}$, achieving the spin flip, that is,
\beq
\label{conditionphase}
\phi_{\sigma}(t_f)=-\frac{m}{\hbar}\int_0^{t_f} \dot{a}_{c}(\tau)x_{0}(\tau)\,d\tau = \frac{\pi}{2},
\eeq
makes the spin rotate around the $z$-axis by the $\pi$-angle.
By combining all the conditions, (\ref{conditionac1})-(\ref{conditionphase}), we solve the polynomial ansatz, $a_{c}(t)=\sum^{6}_{j=0}c_{j}t^{j}$,
and obtain
\begin{equation}
a_{c}(t)=-{231\pi}\frac{\hbar}{md}\frac{\omega^{2}t_f^2}{5\omega^{2}t_{f}^{2}-66}\left(s^{6}- 3 s^{5} +3 s^{4} - s^{3}\right).
\end{equation}
Once $a_{c} (t)$ is fixed, we calculate $\alpha (t)$ from Eq. (\ref{eqact}).

\begin{figure}[]
\scalebox{0.58}[0.58]{\includegraphics{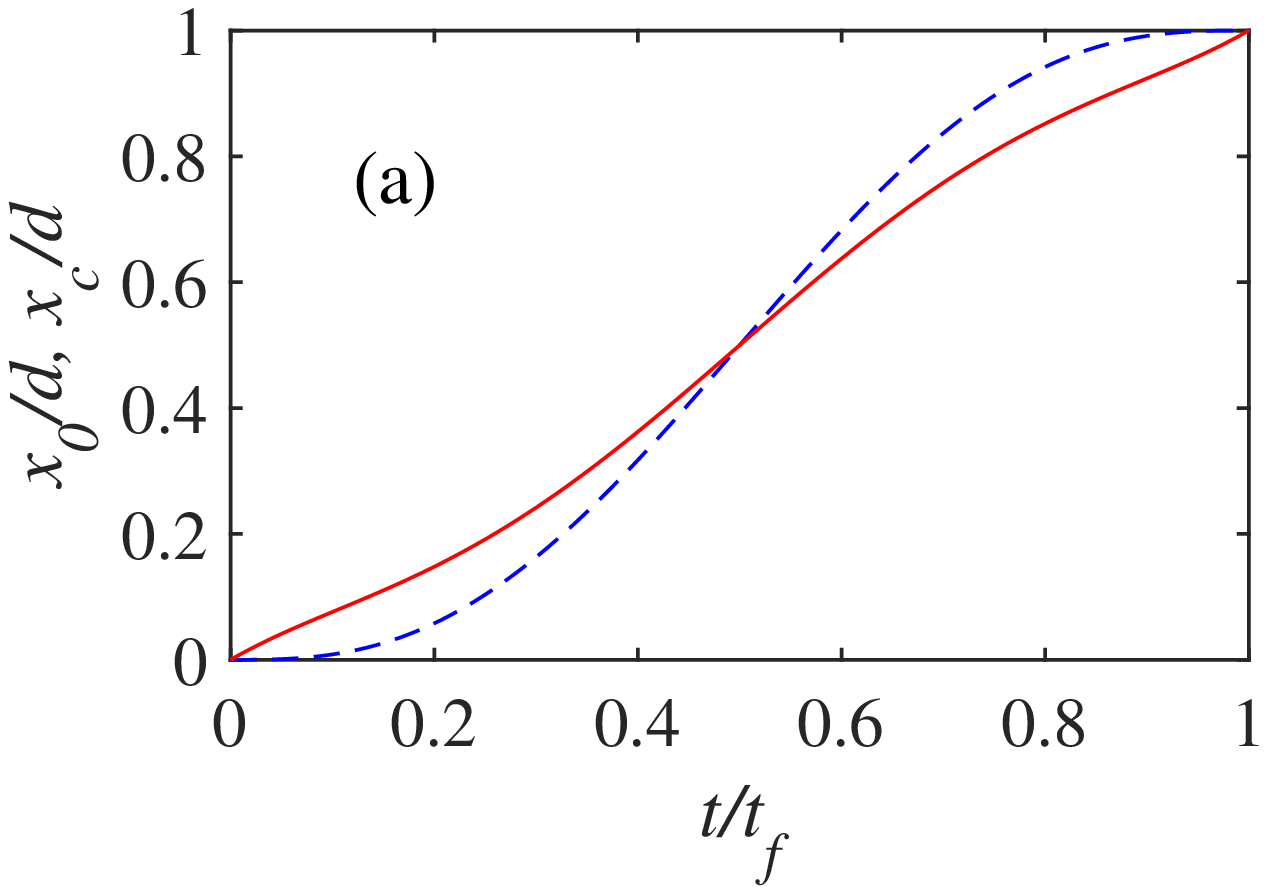}}
\\
\scalebox{0.58}[0.58]{\includegraphics{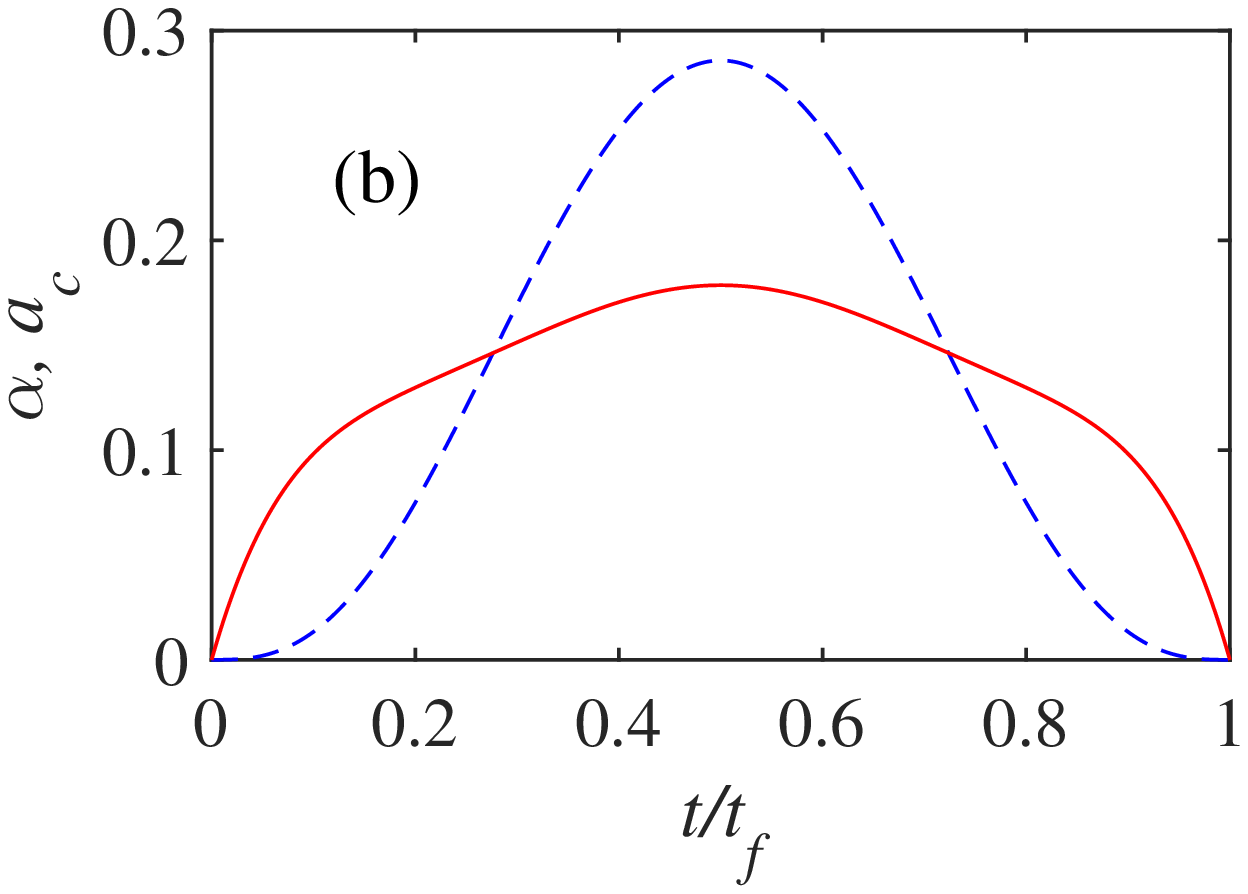}}
\caption{(a) Dependence of the minimum position $x_{0}/d$ (solid red line) and
the center-of-mass $x_{c}/d$ (dashed blue line) on time $t/t_f$.
(b) Dependence of the SOC strength $\alpha$ (solid red line) and the parameter $a_{c} $ (dashed blue line) on time $t/t_f$.
Parameters: $t_f=8/\omega$ and $d=10 $.}
\label{trajectory}
\end{figure}

In a realistic setup, the SOC was realized for $^{87}\mbox{Rb}$ atoms in an external potential, where
the mass of atom is $m=1.443 \times 10^{-22}$ g and the confining potential frequency is $\omega=2\pi \times 250$ Hz.
To simplify the numerical calculations below, we choose $m=\hbar=\omega=1$ and
the units of relevant physical parameters are re-scaled by $T=1/\omega \approx 0.637$ ms
and the characteristic length $a_0=\sqrt{\hbar/(m\omega)} \approx 0.682$ $\mu$m, correspondingly.
Figure \ref{trajectory} demonstrates the designed position $x_0(t)$ and time-dependent SOC strength $\alpha(t)$.

\section{Fast transport and spin dynamics}

\begin{figure}[tbp]
\centering
\scalebox{0.58}[0.58]{\includegraphics{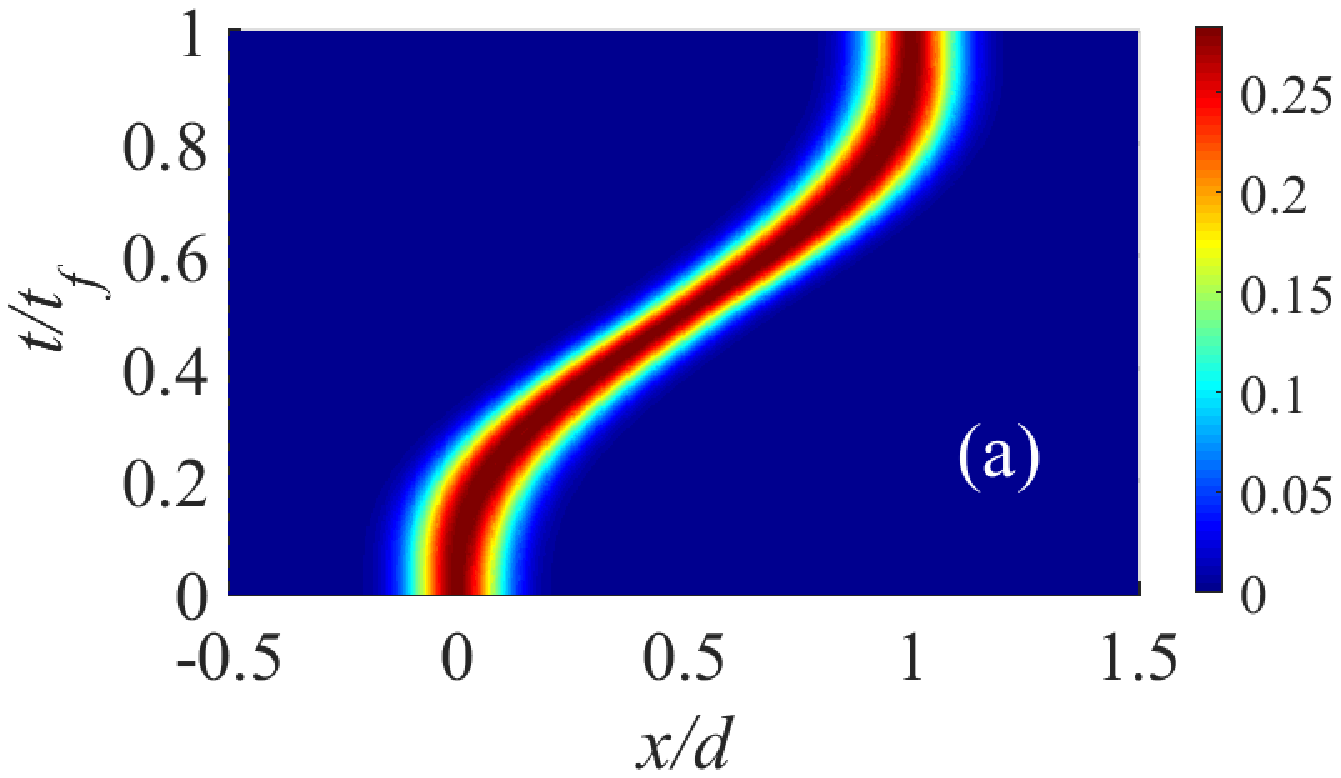}}
\\
\scalebox{0.58}[0.58]{\includegraphics{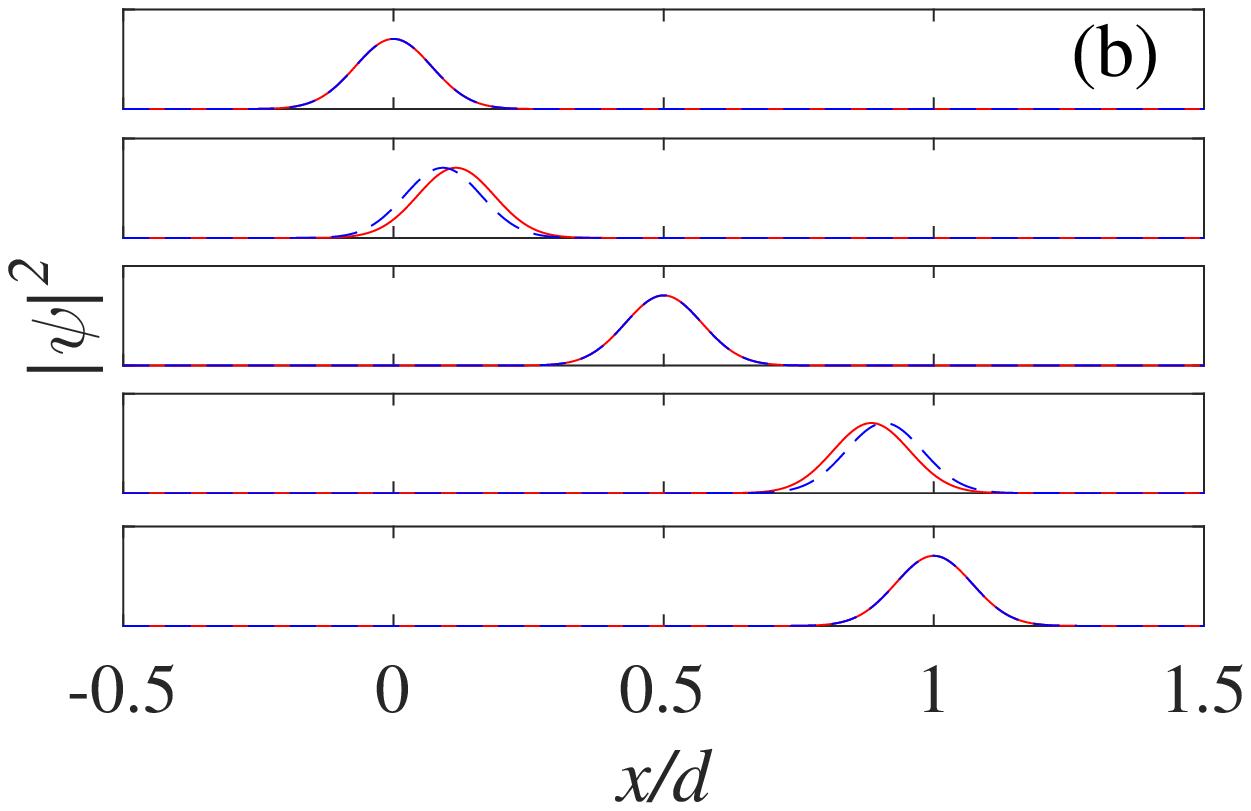}}
\caption{(a) Contour plot of the wave packet propagation during the fast transport designed by the inverse
engineering method. (b) Time evolution of the wave packet with spin-up (solid red line)
and spin-down (dashed blue line) components at different times: $t=0, t_f/4, t_f/2, 3 t_f/4, t_f$.
Parameters are the same as those in Fig. \ref{trajectory}. }
\label{transport}
\end{figure}

For the sake of simplicity, we first consider the fast transport and spin flip without interatomic interaction. We assume the initially prepared state,
\beq
\label{initial}
|\Psi(x,0)\rangle =\frac{1}{\sqrt 2} {1\choose 1}\otimes |\psi (x,0)\rangle,
\eeq
where
\beq
|\psi (x,0) \rangle=\left(\frac{1}{\pi a^2}\right)^{1/4}\exp{\left[-\frac{x^2}{2 a^2}\right]}
\eeq
is the ground state in the harmonic potential, which we consider here without loss of generality. The
final wave function has the form
\beq
\label{final}
|\Psi(x,t_f)\rangle = \frac{1}{\sqrt 2} {~1\choose -1} \otimes |\psi (x,t_f)\rangle,
\eeq
with
\beq
|\psi (x,t_f) \rangle =\left(\frac{1}{\pi a^2}\right)^{1/4}\exp{\left[-\frac{(x-d)^2}{2 a^2}\right]}.
\eeq
\begin{figure}[]
\scalebox{0.58}[0.58]{\includegraphics{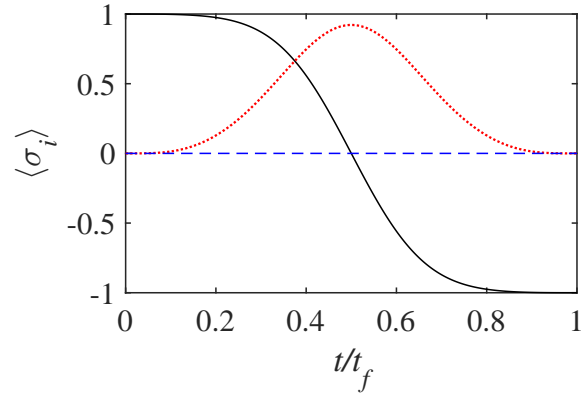}}
\caption{Time evolution of spin components $\langle \sigma_i \rangle $ during
the fast transport representing $\langle \sigma_x \rangle$ (solid black line), $\langle \sigma_y \rangle$ (dotted red line),
and $\langle \sigma_z \rangle$ (dashed blue line). 
Parameters are the same as those in Fig. \ref{trajectory}.}
\label{spin}
\end{figure}

Figure \ref{transport} (a) demonstrates that by using the designed trap position,
the BECs is transported from $x_{0}=0$ to $x_{0}=d$ without any final excitation.
To illustrate the spin motion, the propagation of spin components is
displayed in Fig. \ref{transport} (b).
Since the initial spin parallel to the $x$-axis is not an eigenstate of the Hamiltonian (\ref{Hamiltonian}),
the spin starts to rotate and the wave packet splits into two components having different velocities.
To understand this splitting,
we define the velocity operator, taking into account the spin-dependent contribution,
\begin{equation}
v=\frac{i}{\hbar}[H,x]=\frac{p}{m}+\alpha \sigma_z.
\label{velocity}
\end{equation}
As a result, the SOC leads to different velocities for  spin-projected components of wave packet.
Initially, the spin is parallel to the
$x$-axis, the expectation value of the velocity vanishes, $\langle v \rangle= 0$ at $t=0$, the two components of the
wave packet coincide and start to split.
At the middle time $t=t_{f}/2$, the spin is parallel to the $y$-direction, and
the two components merge again with the same but non-zero expectation value of velocity.
At the final time, based on the boundary conditions, (\ref{conditionac1})-(\ref{conditionac3}),
the spin is antiparallel to the $x$-direction, such that the spin
components of the wave packet coincide and $\langle v \rangle= 0$, see Fig. \ref{transport} (b).

Next, we discuss the spin evolution in terms of the reduced density matrix \cite{Zhenya}
\begin{equation}
\rho(t)=|\Psi (x,t)\rangle \langle \Psi (x,t)|=\left[
\begin{array}{cc}
  \rho_{11}(t) & \rho_{12}(t) \\
  \rho_{21}(t) & \rho_{22}(t) \\
\end{array}\right],
\end{equation}
where 
$$
  \rho_{ij}(t) = \int \Psi_i(x,t)\Psi_j^*(x,t)\,dx, ~(i,j=\uparrow, \downarrow),
$$
and ${\rm tr}(\rho) = \rho_{11}(t)+\rho_{22}(t)=1$ because of the normalization of the wave function. 
As a consequence, the three spin components can be defined by $\langle \sigma_i \rangle =$ ${\rm tr}(\sigma_i \rho)$ ($i=x,y,z$).
Figure \ref{spin} shows the time evolution of the spin components, where the expectation value of
spin polarization at initial time $t=0$ is $\langle \sigma_x \rangle(0)=1$, and $\langle \sigma_x \rangle(t_f)=-1$ at the final time $t=t_f$.
As a result, the spin flips with rotation around the $z$-axis, when the BECs is being transported.
Additionally, we define the length of the spin vector inside the Bloch sphere as $P=(\sum_{i=1}^{3}\langle\sigma_i\rangle^2)^{1/2}$. At
the initial and final times $P=1$, implying that
the spin is in a pure state on the Bloch sphere. During the non-adiabatic transport, spin-dependent excitations of the orbital modes occur,
resulting in a mixed state in the spin subspace with $P<1$.  

\section{Comparison with adiabatic transport with constant SOC}

For comparison,  we now consider the case of a constant SOC strength, $\alpha$, and the adiabatic transport with a constant velocity, $d/t_f$.
The adiabatic transport for linear protocol \cite{ZhangQ} requires $t_{f}^{\rm ad}\gg d\sqrt{m/(2\hbar \omega)} \approx 7.07,$ where we take $d=10$
as in Fig. \ref{trajectory}.
In the adiabatic limit, neglect the derivatives $\ddot{x}_c$ and $\dot{x}_c$, resulting in
$x_c(t) =x(t)$. Substituting the constant SOC strength $\alpha$, we solve Eq. (\ref{eqact}) and
finally obtain
\beq
a_c (t) = \alpha[1- \cos(\omega t)],
\eeq
with the initial boundary conditions, $a_c (0)= \dot{a}_c (0) =0$. Obviously, when $\omega t_f = 2 \pi k $ ($k=1,2,3,\ldots$), the boundary conditions
$a_c (t_f) = \dot{a}_c (t_f) =0 $ are fulfilled.
On the contrary, the other solution $a_c= \alpha$ is neglected, since in this case the spin part of unitary transformation $U(t)$ becomes $U_s= e^{-i m \alpha x \sigma_{z}/\hbar}$, and the initial and final spin states
are transformed, due to $U_s(0)=U_s(t_f) \neq 1$.
Furthermore, the phase factor (\ref{phase}) is calculated as
\beq
\phi_{\sigma}(t_f)= -\frac{m \alpha d}{\hbar \omega t_f} [\sin(\omega t_f) - \omega t_f  \cos(\omega t_f)],
\eeq
by using $\dot{a}_c = \alpha \omega \sin(\omega t)$ and $x_0(t) = d t/t_f$.
When $\omega t_f = 2 \pi k $ ($k=1,2,3,\ldots$), such phase factor is simplified as
$\phi_{\sigma}(t_f) = d / \lambda_{\rm so}$ with $\lambda_{\rm so} = \hbar/ (m \alpha)$. As a consequence,
we see from Eq. (\ref{U}) that the spin is rotated by the angle $2  d /\lambda_{\rm so} $ around
the $z$-direction. By further imposing $\phi_{\sigma}(t_f) = \pi/2,$
we have the characteristic length for spin flip
\beq
d_{\rm sp}= \pi \lambda_{\rm so}/2,
\eeq
and the spin-flip time is $t_{\rm sp}=  (d_{\rm sp}/ d) t_f$. This is consistent with the result of Ref. \cite{RamsakNJP}.

\begin{figure}[]
\scalebox{0.58}[0.58]{\includegraphics{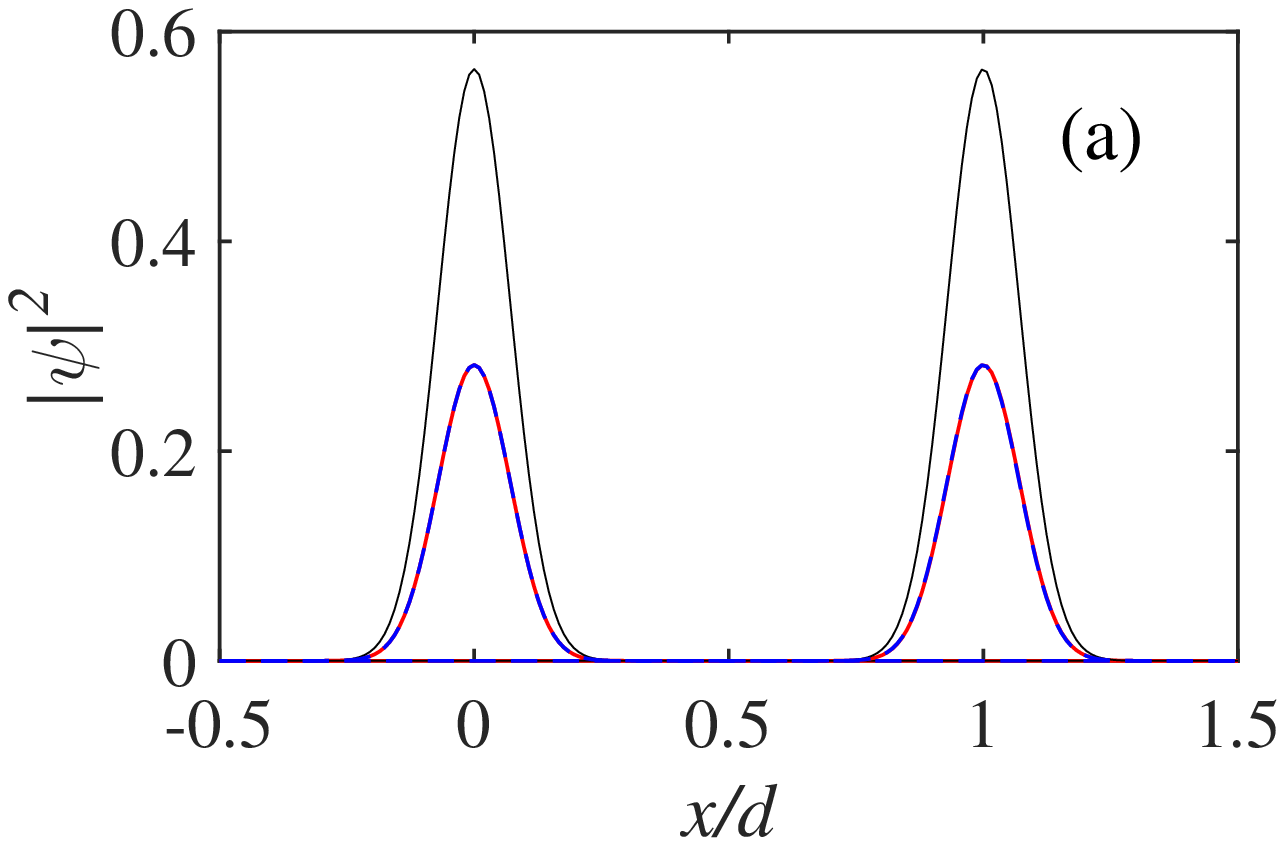}}
\\
\scalebox{0.58}[0.58]{\includegraphics{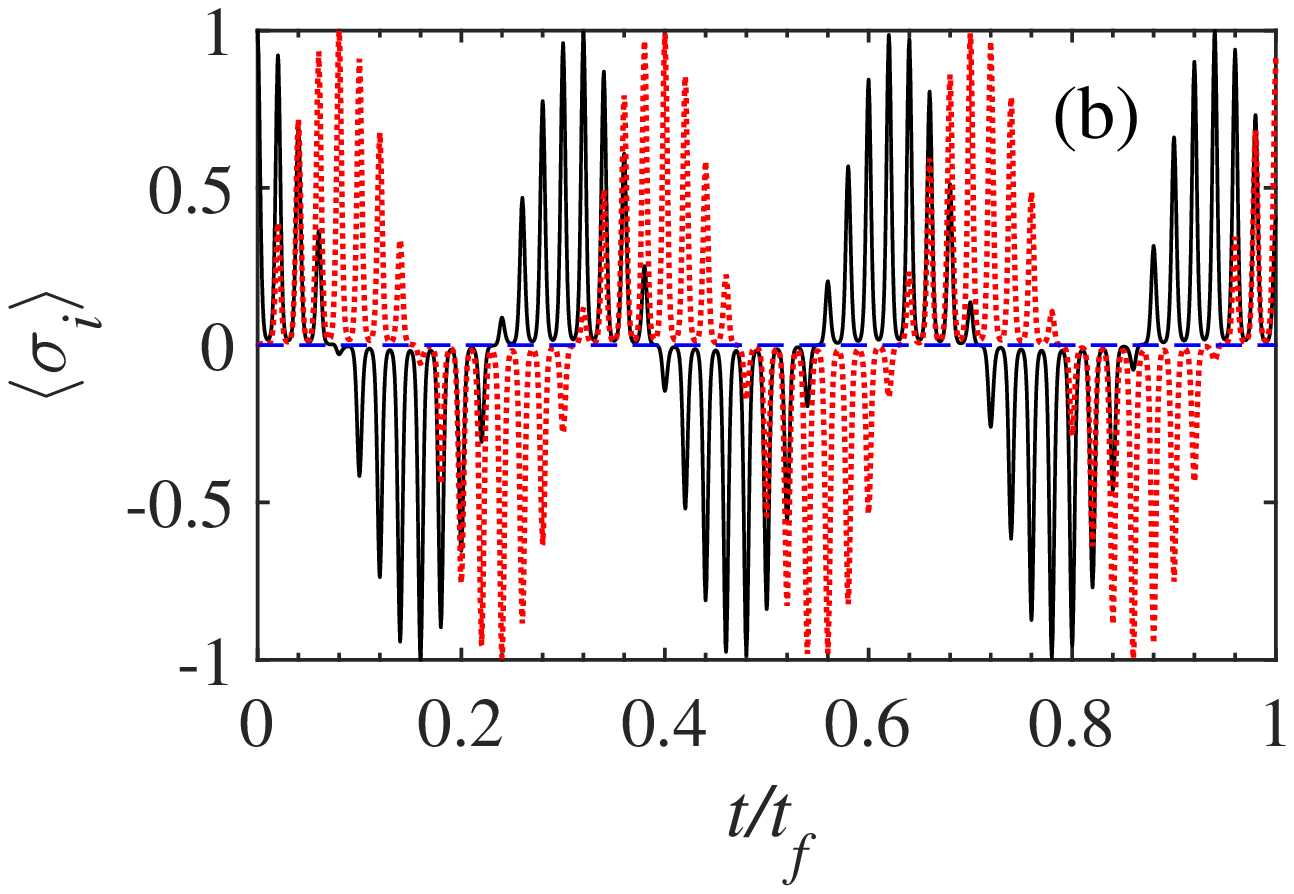}}
\caption{(a) Profiles of total density $|\Psi(x,t)|^2$ (solid black line) and density of spin components,
$|\Psi_{\uparrow,\downarrow}(x,t)|^2$ (solid red and dashed blue lines, undistinguishable) at the initial
time $t=0$ and at the final time $t=t_f$. (b) Time evolution of spin components $\langle \sigma_i \rangle $
during the fast transport  representing
$\langle \sigma_x \rangle$ (solid black line), $\langle \sigma_y \rangle$ (dotted red line),
and $\langle \sigma_z \rangle$ (dashed blue line). Parameters: $\alpha=1, t_f=100 \pi/\omega, d=10$.}
\label{linearwave}
\end{figure}

Figure \ref{linearwave} (a) illustrates that the BECs can be transported from $x_{0}=0$ to $x_{0}=d$
when $t_f = 100 \pi/ \omega$ is sufficiently long to satisfy the adiabatic criteria.
When the time $t_f$ is an integer multiple of $2 \pi/\omega$, there is no final excitation of the orbital motion, due to $a_c (t_f) = \dot{a}_c (t_f) =0$ in Eq. (\ref{U}).
The orbital wave function is exactly displaced by $d$ in this case.
Meanwhile, the spin dynamics is determined by the characteristic length $d_{\rm sp}$. In other words,
the spin flip can be realized if the transported distance is $d_{\rm sp}$.
When $d=10$ and $\alpha=1$, we can obtain the periodical time $t_{\rm sp}=0.157 t_f$ for spin flip,
see Fig. \ref{linearwave} (b). In this case, the final spin state is not the eigenstate of $\sigma_{x},$ and
the spin cannot be flipped completely, since $d \neq (2k-1) d_{\rm sp}$ ($k=1,2,3,\ldots$). However, at the final time
the wave functions of the spin components coincide being characterized by the same displacement $d$, see Fig. \ref{linearwave} (a),
since the terms on $a_c$ and $\dot{a}_c$ in Eq. (\ref{U}) vanish at $t=t_f$.
Therefore, one can transport atoms from $x_{0}=0$ to $x_{0}=(2k-1)d_{\rm sp}$ adiabatically with flipping the spin simultaneously.
Note that in the adiabatic approximation, the characteristic spin rotation length depends only on the SOC strength being independent of
the transport velocity. As compared to the inversely designed protocols, the adiabatic transport with constant SOC and velocity can achieve the same effect on the
orbital and spin motion, but it requires a much longer time and possible only for a relatively small set of final positions.

\section{The effects of interatomic interactions}

In this section, we briefly present the influence of interatomic interaction on the orbital
motion and spin dynamics designed by the inverse engineering method.
The Hamiltonian is rewritten as
\beq
\mathcal{H}=\frac{p^{2}}{2m}+\frac{1}{2}m\omega^{2}[x-x_{0}(t)]^{2}+\alpha(t)p\sigma_{z}+g|\Psi(x,t)|^2,
\eeq
where the repulsive interaction characterized by parameter $g=2 a_s\hbar\omega_{\perp}>0$ is involved, with the scattering length $a_s$ and the transverse
confinement frequency $\omega_{\perp}\gg \omega$. The nonlinearity $g$ can experimentally be adjusted by Feshbach resonances and the transversal confinement.
In the following numerical calculations, the dimensionless $\tilde{g} = 2 \tilde{a}_s \tilde{\omega}_{\perp}$ with $\hbar =1$,
where $\tilde{a}_s$ is the scattering length in the units of $a_0$ and $\tilde{\omega}_{\perp}$ is the transverse confinement frequency in
the units of $\omega$. Here $\Psi(x,t)$ is the wave function of the condensate described by the mean-field Gross-Pitaevskii (GP) equation, and its normalization is
$\int^{+\infty}_{-\infty} |\Psi(x,t)|^2 dx = N$, with the number of atoms $N$.
In general, one can transport the ground state of the Gross-Pitaevskii (GP) equation as the initial wave packet \cite{Zhenya}.
Instead, for consistency we assume the following initial Gaussian wave packet
\beq
\label{initialnonlinear}
|\Psi(x,0) \rangle=\frac{1}{\sqrt 2} {1\choose 1}\otimes \left(\frac{N^2}{\pi a^2}\right)^{1/4}\exp{\left(-\frac{x^2}{2 a^2}\right)}.
\eeq
This Gaussian assumption turns out to be an appropriate choice, particularly for weak interaction $g$ and small number of atoms $N$ \cite{Zoller,Haque},
otherwise the wave function becomes an inverted parabolic shape at strong repulsive interaction, especially $N>600$, see \cite{Sakhel}.

\begin{figure}[]
\scalebox{0.58}[0.58]{\includegraphics{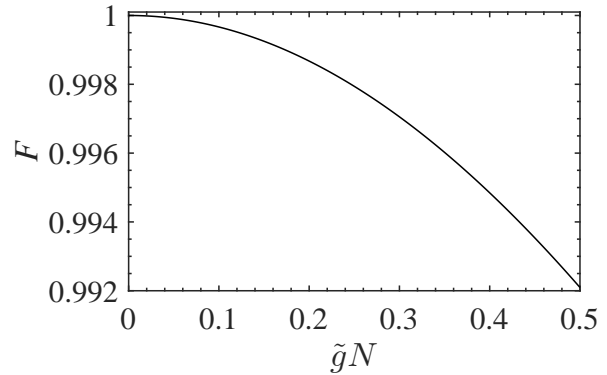}}
\caption{Fidelity versus $\tilde{g}N$, resulting from interatomic interaction, where $N=100$ and other parameters are the same as those in Fig. $\ref{trajectory}$.}
\label{fvsg}
\end{figure}

The influence of the interatomic interaction and the numbers of atom is illustrated by the fidelity,
$F=|\langle \tilde{\Psi}(t_f)|\Psi(t_f)\rangle|^2$, see Fig. \ref{fvsg}, where the target state is defined as
$$
| \tilde{\Psi}(t_f) \rangle=\frac{1}{\sqrt 2} {~~1\choose -1}\otimes \left(\frac{N^2}{\pi a^2}\right)^{1/4}\exp{\left[-\frac{(x-d)^2}{2 a^2}\right]},
$$
with the displacement $d$ and spin flip, and the wave function $|\Psi(t_f)\rangle$ is the numerical result
calculated by the split operator method.
Figure \ref{fvsg} demonstrates that the Gaussian approximation is good to describe such a non-linear system, especially when $\tilde{g} N$ is reasonably smaller than $1$.
However, the fidelity becomes worse with increasing the nonlinearity $g$ and the number of atoms $N$.
To understand this effect, we shall analyze the time evolution of two spin components.
As mentioned above, due to the SOC the initial wave packet starts to split in two spin components with different velocities.
The repulsive interaction helps separation and hinders merging. The spin dynamics and the orbital motion become different
from the linear case ($g=0$), and the two spin components cannot merge at $t= t_f/2$ and $t=t_f$.
This causes the final separation of the spin components and decreases the fidelity.


\section{CONCLUSION}

We have presented a method for achieving the fast transport and spin control of spin-orbit coupled BECs
in moving harmonic potentials. The inverse engineering, based on the concept of ''shortcuts to adiabaticity``
is applied to design the potential position and the time-dependent strength of SOC, by choosing
appropriate boundary conditions. The adiabatic transport with a constant SOC has been compared with the developed protocol
to illustrate the advantage of the shortcut-based design. Finally, we have discussed the SOC BECs transport taking into account the interatomic interaction at the level of the
Gross-Pitaevskii equation.

The inverse engineering method proposed here is helpful for manipulating the SOC BECs and controlling spin-orbit coupled qubits by designing
the time-dependent SOC and the potential motion. This might have applications in quantum information processing, atom interferometry,
and quantum metrology. Several natural extensions of this approach can be done in the near future.
For instance, one can hybridize the inverse engineering and the optimal control theory \cite{Lu14,David2} to optimize the shortcuts in the presence of noise
and device-related errors. Being combined with the variational principle \cite{Zoller} and hydrodynamic approach \cite{Stringari}, the shortcuts can be further designed for soliton dynamics \cite{Jing} or quench dynamics in the SOC BECs.
Last but not least, our system resembles electron confined in parabolic quantum dots or wires \cite{ZhenyaPRB,JQYouPRL,RamsakNJP,RamsakPRL,RamsakPRL},
which can be useful for generating spin-dependent coherent and Schr\"{o}dinger cat states \cite{Schrodingercat}.

\section*{ACKNOWLEDGMENTS}

We thank Yongping Zhang, Thomas Busch, and Guanzhuo Yang for helpful discussions.
This work is partially supported by the NSFC (11474193, 61404079), the Shuguang (14SG35),
and the Program for Professor of Special Appointment (Eastern Scholar).
E.Y.S. acknowledges support of the University of the Basque Country UPV/EHU
under program UFI 11/55, Spanish MEC/FEDER (FIS2015-67161-P) and Grupos
Consolidados UPV/EHU del Gobierno Vasco (IT-472-10). Y.B. also acknowledges Juan de la Cierva program.

\end{document}